\documentclass[sigconf]{acmart}

\usepackage{subcaption}
\usepackage{float}

\AtBeginDocument{%
  \providecommand\BibTeX{{%
    \normalfont B\kern-0.5em{\scshape i\kern-0.25em b}\kern-0.8em\TeX}}}

\acmConference[SIGSPATIAL ’21]{SIGSPATIAL ’21}{November 02–-05, 2021}{Beijing, China}
\setcopyright{none} 

\begin{document}

\title{Spatial Data Mining of Public Transport Incidents reported in Social Media}

\author{Kamil Raczycki, Marcin Szymański,\\Yahor Yeliseyenka}
\authornote{The first three authors contributed equally to this research.}
\email{raczyckikamil@gmail.com,}
\email{{szymanski7marcin,yahor.yeliseyenka}@gmail.com}

\affiliation{%
  \institution{Department of Artificial Intelligence,\\Wrocław University of Science and Technology}
  \city{Wrocław}
  \country{Poland}}

\author{Piotr Szymański, Tomasz Kajdanowicz}
\email{{piotr.szymanski,tomasz.kajdanowicz}@pwr.edu.pl}

\affiliation{%
  \institution{Department of Artificial Intelligence,\\Wrocław University of Science and Technology}
  \city{Wrocław}
  \country{Poland}}

\renewcommand{\shortauthors}{Raczycki et al.}

\begin{abstract}
Public transport agencies use social media as an essential tool for communicating mobility incidents to passengers. However, while the short term, day-to-day information about transport phenomena is usually posted in social media with low latency, its availability is short term as the content is rarely made an aggregated form. Social media communication of transport phenomena usually lacks GIS annotations as most social media platforms do not allow attaching non-POI GPS coordinates to posts. As a result, the analysis of transport phenomena information is minimal. We collected three years of social media posts of a polish public transport company with user comments. Through exploration, we infer a six-class transport information typology. We successfully build an information type classifier for social media posts, detect stop names in posts, and relate them to GPS coordinates, obtaining a spatial understanding of long-term aggregated phenomena. We show that our approach enables citizen science and use it to analyze the impact of three years of infrastructure incidents on passenger mobility, and the sentiment and reaction scale towards each of the events. All these results are achieved for Polish, an under-resourced language when it comes to spatial language understanding, especially in social media contexts. To improve the situation, we released two of our annotated data sets: social media posts with incident type labels and matched stop names and social media comments with the annotated sentiment. We also opensource the experimental codebase. 
\end{abstract}




\maketitle

\section{Introduction}
Public transport companies have embraced social media in the recent decade to perform multiple communicative functions. This rise of communication concerning spatial phenomena has been followed by releases of public transport schedules in standardized GTFS formats across the world. As social media communication is very short-term oriented, often it is up to citizen data science to retrieve information across longer periods of time, aggregate and analyze them to and make wider communities, affected by transport phenomena, aware of long-term changes in how public transport operates. These citizen science challenges affect multiple areas of spatial language understanding, geographic information mining, and aligning language to transport network structures and more sophisticated data sets.

One such case can be made for a Wrocław, where following two decades of under-investment, the local public transport service quality is in critical condition as the city public transport operator - MPK Wrocław - struggles to maintain a 10-day streak without tram derailment. As service disruptions occur with high frequency, MPK Wrocław provides information about anomalies in the trips, using information boards, dedicated applications, and social media (Facebook and Twitter). MPK Wrocław carries about 200 million passengers a year, about 550 thousand a day. It's capacity to inform about service outages is not only critical to maintaining a standard of life quality in daily activities in the city but also indirectly concerns political decisions, mobility behavior, carbon emissions, and many other areas of citizen science activity. 

While day-to-day communication is maintained regularly with low latency, it is never released in aggregated form. More and more often agencies remove old social media posts, to avoid citizens reacting to comments made under outdated content. The grass-root ability to improve local quality of life depends on inhabitants being able to make data-driven arguments. Citizen science therefore, heavily relies on the automatic spatial understanding of social media information related to public transport phenomena. Without an aggregated data set released in an open data fashion, it is crucial to pinpoint where traffic disruption happens in both GIS and mobility network perspectives. Citizens need to be able to map social posts concerning service disruptions to a GPS coordinate and a relevant stop or lines in the transportation network.

We present a method and a study of a public transport operator's service disruption communication in the situation of a persistent crisis. We analyze posts and comments from the official social media channels of MPK Wrocław concerning traffic obstructions and contribute the following results:
\begin{itemize}
    \item a systematic typology of communique and discuss the magnitude of social response,
    \item an approach for classification and spatial understanding of service disruption alerts,
    \item use cases to analyze the aggregated impact on the transportation network and mobility and automatic examination of the sentiment of citizens' feedback.
\end{itemize}

We built a robust method for automatic spatial language processing to allow processing social media communication into a spatial database and show its use cases. Our work concerns the Polish language, which is an under-resourced language in this regard. Our method can work with any source of social media posts, provided a list of stop names mapped to GPS coordinates, or open street map data for the evaluated location. 

\section{Background}

\subsection{How do public traffic operators handle social media?}

\citet{bregman2012uses} discusses five ways traffic operators engage in social media communication activities:

\begin{enumerate}
    \item sharing information about real-time service changes and advisories (\textbf{timely updates}),
    \item providing information about services, fares, and long-range planning projects (\textbf{public information}),
    \item taking advantage of the interactive aspects of social media to connect with their customers in an informal way (\textbf{citizen engagement}),
    \item recognizing current workers and recruiting new employees (\textbf{employee recognition}), 
    \item displaying a personal touch and entertaining their riders pleasurable brand-related content (\textbf{entertainment}).
\end{enumerate}

Bregman finds that 77\% of surveyed transit agencies used Twitter and 49\% used Facebook (multiple answers allowed) to inform about real-time service alerts and 66\% of agencies updated their relevant channels more than once a day with a service update. \citet{lui2016understanding} designed an online survey of the top 50 transport agencies around the USA to collect general information about their purpose and social media usage. They show that public transport agencies are extremely interested in measuring their social media success: more than half of public transport agencies currently measure social media outcomes through a number of “friends and followers”, “likes” and “retweets”. Only a few of them measure their social media program effectiveness with positive perception similar to sentiment analysis. \citet{jeffrey2016management} describes how the Washington State DOT (WSDOT) was chosen as an example of well-executed social media communication capable of limiting uncertainty in high disaster situations and proving to be a recognizable source of reliable information.

\citet{doi:10.1080/01441647.2014.915442} propose 6 stages of the disruption and recovery process on the role of information delivery. Authors state that social media can be a very efficient tool to inform passengers about disruptions but it can exclude a certain group of society that doesn't use smartphones. 
In another review on \cite{pender2014social} present how transit agencies use social media during unplanned transit network disruptions. \citet{GALTZUR2014115} focuses on categorizing live feed of tweets on an example of 3 mln tweets tweeded during and before Liverpool soccer matches. Using bag-of-words models, authors propose 2 models: one to recognize if the tweet was tweeted by an authority (city, transport company) and second to recognize if the tweet was about transport. Authors propose a hierarchical approach to categorize tweets further in 3 main branches: information about transport needs, detection of irregular events that impact on mobility and traveler opinions about transport services.

Further studies include evaluating how transit agencies communicated during large events \cite{cottrill2017tweeting} or how crowd sourced data can complement agency communication of disruptions \cite{congosto2015microbloggers}. An important area of contemporary social media and public transport research includes using social media check-ins for mobility research \cite{lv2017social, tu2017coupling, rashidi2017exploring,serna2017sustainability},  planning decisions \cite{ni2016forecasting, nikolaidou2018utilizing}, transport performance \cite{haghighi2018using}, user activities \cite{chaniotakis2017inferring} and satisfaction \cite{alshehri2019analyzing}. Recently new technologies emerged to push service disruption information such as bots and chat bots \cite{gault2019travelbot}.

\subsection{Social media meets citizen science}

Social media profiles of public transport agencies are also important for citizen science. \citet{perriam2019} describes how non-academic organisations can use data from social media on the example of the public sector organization "Transport for London". He focuses on the ethnographic research into the workflow of the organisation and shows that TfL uses social media (here tweets in particular) not only for customer service but also for staff management, productivity observation and decision making.

\citet{li2019mining} discuss how one can collect public opinions from social media to extract from them potential improvements proposals or use them as a support in decision making. The researchers describe the case of the subway in the city of Nanjing in China using the data collected from the Sina microblog. Three types of analysis have been made (sentiment analysis, spatiotemporal analysis and topic/subject analysis) to localize the most important topics raised in public space. They note that passengers comment mainly on the Metro congestion, air-conditioning temperature, environment in the carriage, equipment failure, and so on. Other comments have been summed up to formulate some example prescriptions: what should be improved/done according to citizens.

\citet{ragothama2016analytics} describe the analysis of delays using the data (mainly GTFS data, besides that for example timetables) from the cities of Rome and Stockholm. For Stockholm, the focus was placed primarily on recognizing sources and factors of delays, for Rome the ability to predict future delays was more important. Generally, Ragothama et al. presented the process of working with geospatial and real-time data and they pointed out potential issues in such analysis.

\section{Data collection}
MPK Wrocław publishes information about impediments and changes in a public transport schedule using a variety of available communication channels: a dedicated mobile app - iMPK; timetable's display boards located near bus and tram stops; social media - official Facebook and Twitter profiles.

\begin{figure}[h!]
  \centering
  \includegraphics[width=\columnwidth]{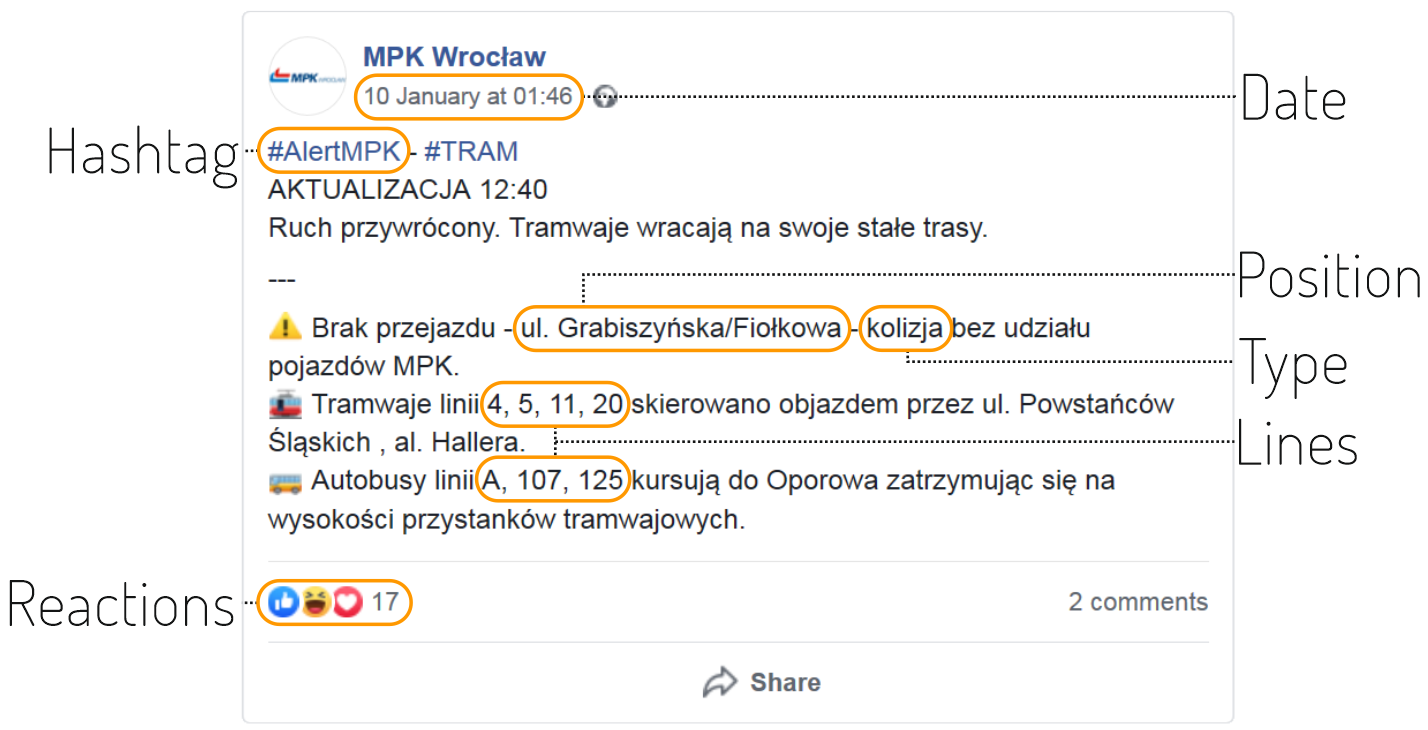}
  \caption{Example Facebook post with highlighted information extracted from it.}
  \label{fig:example}
\end{figure}

Between 2017 and 2019 we collected 4161 tweets, 1680 Facebook posts with MPK communique concerning various topics of public transport operations. Additional metadata from social media was acquired from the content's publication date and reactions, including 24809 comments collected. Because MPK Wrocław posts diverse messages on Facebook, hashtag \textbf{\#AlertMPK} was used to filter posts that reported incidents and other information concerning schedule changes. Figure \ref{fig:example} shows an example of a post and data acquired from it. After filtering there were 622 posts left for an analysis\footnote{We were unable to assure the deanonymization of our GitHub repository, however in the final version of the paper we will provide a link to the repository with relevant code, data sets, trained models and a streamlit dashboard for exploring current twitter data stream.}. 

These textual data points contain spatial information we were interested in extracting, the two main aspects  We were interested in obtaining the spatial understanding of \textbf{incident locations} and being able to relate them to passenger mobility characteristics. We obtained bus and tram stops positions as well as schedules from open data GTFS repositories and mobility information from the recent survey of the metropolitan area mobility \cite{KBR2018}, which divides the Wrocław area into regions and includes SIM-card based approximation of passenger exchange between regions. 

\section{Spatial Understanding}

\begin{figure}[h!]
  \centering
  \includegraphics[width=\columnwidth]{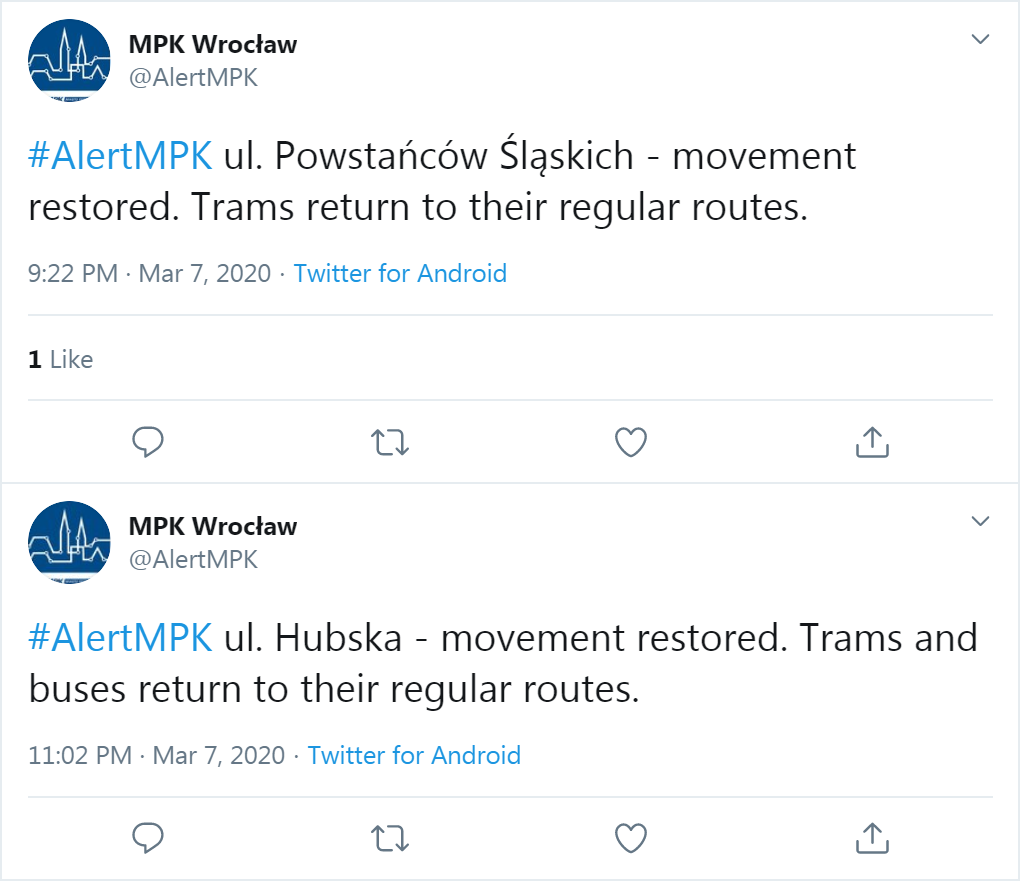}
  \caption{Example tweets of type \textit{fix} - they use the same template, there is only a difference in the street name. Tweets translated by authors.}
  \label{fig:fix}
\end{figure}

One can identify in the posts of MPK the specific language they use. The posts with a similar content type tend to use the same template or to have a similar structure (Figure \ref{fig:fix}). There was additional information derived from acquired raw data: type of incident, the position of the incident, and lines affected by the change in a schedule.

Maintaining a template-based information system allows the public transport operator to decrease the cognitive load needed to understand the post, which is usually received by the user in a stressful context related to service disruption. The structured form also builds the foundation for trust as the users get used to the communication style and identify it with the operator's brand. Such posts are also easier to parse for citizen sciences tasks. Finally, when an irregular situation is reported, its non-template form increases the alertness of the reader.

Following manual exploration we distinguished multiple types of the content in the MPK's posts. As good categorization of posts is essential in the context of future applications we decided to propose the following typology of transport information:

\begin{itemize}
    \item \textbf{accident} - traffic accidents (including accidents without MPK's vehicles)
    \item \textbf{event} - events affecting urban traffic, e.g. matches, marathons, special (e.g. holiday) timetable
    \item \textbf{fix} - information about restoring the state from before the failure
    \item \textbf{incident} - unwanted, random events, e.g. police intervention, fire brigade action, aggressive passenger, medical intervention, etc.
    \item \textbf{malfunction} - failures of infrastructure (e.g. power outages, derailments) and vehicles (e.g. broken pantograph, another vehicle malfunction)
    \item \textbf{renovation} - planned infrastructure repairs
    \item \textbf{unknown} - traffic difficulties with no specified cause
\end{itemize}

We manually tagged 3532 posts with the above classes and build a classifier using the FastText model \cite{joulin2017bag}. 3000 of those were used in a training process and the rest was used for testing. Given the structured language of the posts, even a simple embedding based model was able to achieve the 87\% accuracy for the test set. We use the model to classify the Facebook posts, which are often lengthier but include the structured information as one of the paragraphs. 

\begin{figure}[h]
  \centering
  \includegraphics[width=\columnwidth]{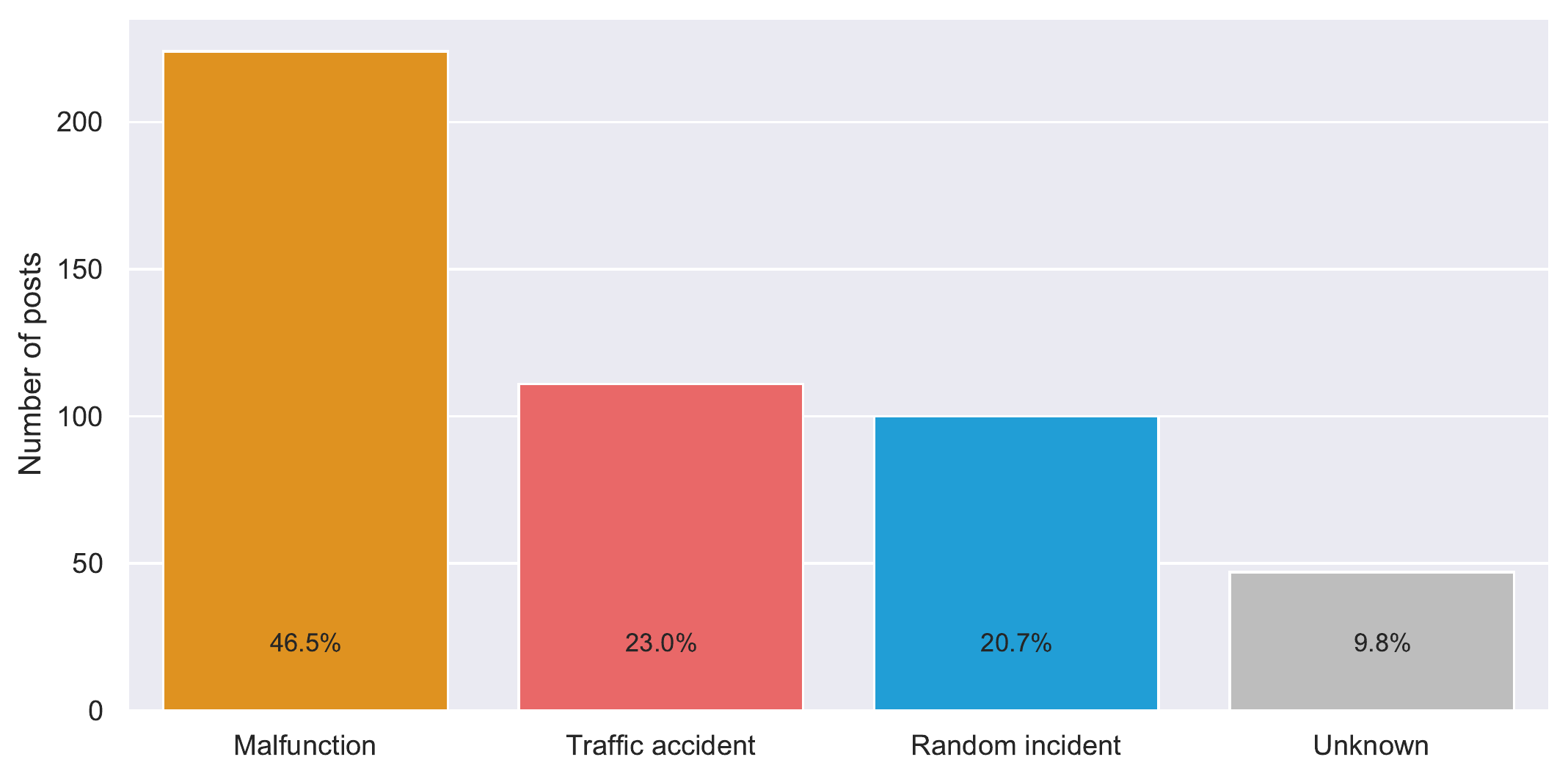}
  \caption{Distribution of classified \#AlertMPK posts.}
  \label{fig:post-types-chart}
\end{figure}

We excluded three of the proposed classes from further analysis: \textbf{event} and \textbf{renovation}, because they were dedicated to planned changes in schedules, not crisis management; and the \textbf{fix} class because it informed about the restoration to normal and not about disruptions. Other classes were further analyzed as unplanned events that caused disruptions in traffic. After filtering there were 482 posts left. Distribution of those posts types can be seen in a figure \ref{fig:post-types-chart}. 

It is important to mention, that although some of the posts (especially Facebook ones) seem to be little equivocal, ambiguous, typically one can identify the correct class quite easily and, as it was measured with Cohen's kappa at 0.75 for each pair of annotators (3 annotators altogether), the multi-class approach turned out to be inferior compared to the single class division approach.

The spatial understanding was performed by rule-based regexp matching, due to the lack of character-level embeddings for the polish language and important impact of inflection on stop and street names in posts. We analyzed word sequences appearing before the introduction or enumeration of spatial terms such as street names or stop names. Next, we matched word sequences that followed these prefixes to the street and stop name lists available in GTFS and OpenStreetMap data, using Levenshtein distance and thresholding. 

Positions of disruptions were obtained using street names mentioned in posts. Official stops positions were used whenever possible. Intersections were found using Google Maps and the rest of the streets were approximated using Open Street Maps. Figure \ref{fig:basic-map} shows map of disruptions with color-coded types. 

\begin{figure}[h]
  \centering
  \includegraphics[width=\columnwidth]{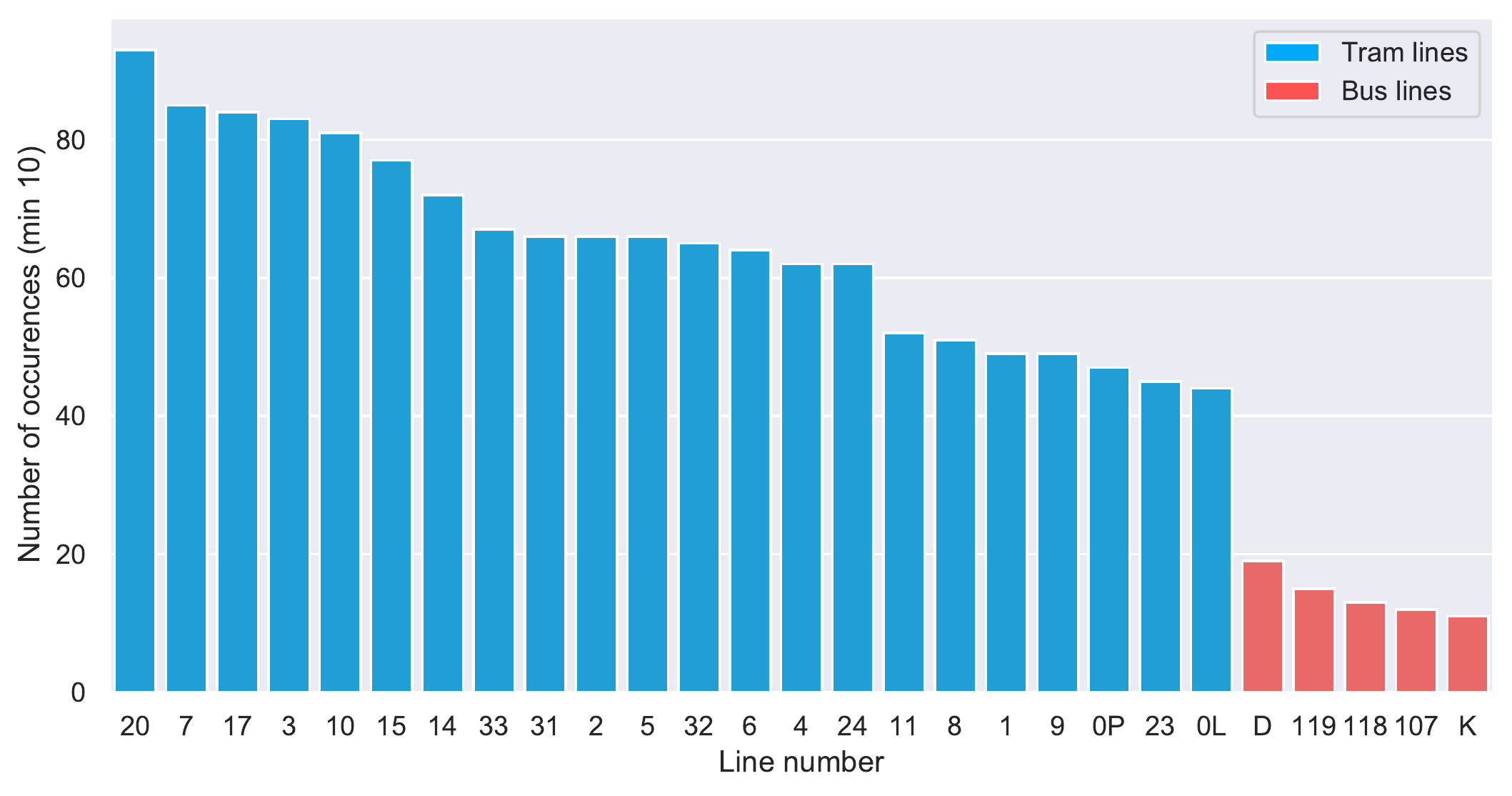}
  \caption{The number of \#AlertMPK posts in which the tram and bus lines were mentioned.}
  \label{fig:lines-mentions-number-chart}
\end{figure}

To detect most affected bus and tram lines, every occurrence of a line's name or a number among different tweets with \#AlertMPK hashtag were found and the number of different tweets in which the line was mentioned was counted. The results are shown in the figure \ref{fig:lines-mentions-number-chart}.

\section{Applications and use cases}

\subsection{Citizen science: understanding the impact on the quality of life}

\begin{figure}[h]
  \centering
  \includegraphics[width=\columnwidth]{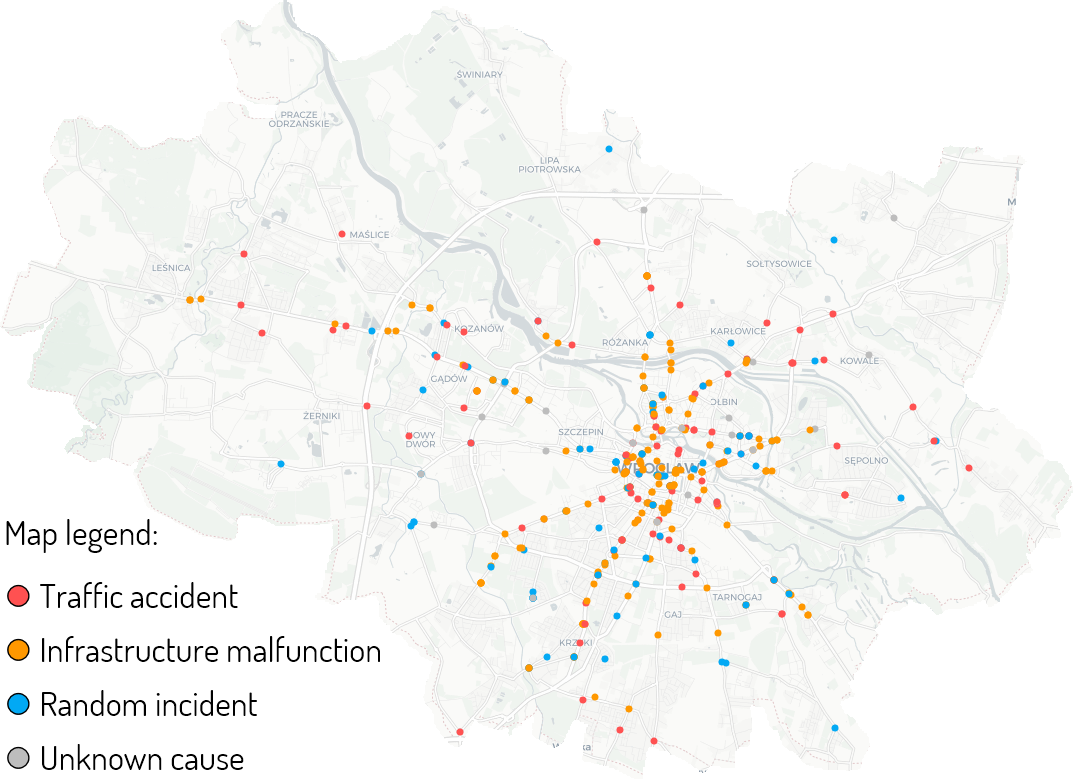}
  \caption{Map of all disruptions from obtained data set with color coded types. }
  \label{fig:basic-map}
\end{figure}

On average there one disruption happened every 2.28 days and almost half of all disruptions were caused by malfunctioning infrastructure or vehicles. As shown in Figure \ref{fig:basic-map} the spatial distribution of incidents spanned the entire tram network in the city and had an important impact on the quality of life. Having extracted the spatial characteristics of each of the incidents we can evaluate how both lines and passengers were affected in more detail.

The most frequently mentioned lines are the tramlines. Even the least mentioned tram lines appeared in tweets much more often than the bus lines. A probable reason for that is that any malfunction or incident related to a tram is far more critical for the traffic infrastructure than the bus's one. A damaged tram blocks the tracks along which other vehicles move, so even an accident of a single tram may cause exclusion of many other trams from traffic or at least change their routes. And the changes are noted in the \#AlertMPK tweets. Buses are more resistant to traffic changes, thus they are mentioned far less frequently.

The perspective on the passenger mobility impact can be drawn from relating the data to the 2018 mobility survey. The survey provides raw data which includes SIM-card based trips counts between regions. As public transport and car mobility are intertwined by the Downs–Thomson paradox \cite{mogridge1987downs} we can treat the passenger mobility data jointly, following the assumption that tram derailment, bus accidents and other public transport incidents negatively impact all means of travel on the affected mobility route and its alternatives in the surround region. We mapped the points of incidents to mobility regions in the study and aggregating the passenger counts for the hour at which the incident occurred. This allows us to estimate that at least 732 618 passengers may have been impacted by service disruptions in the 244 weekdays in which incidents were noted, which is an average of 3000 per day. This is a lower bound estimation as it is calculated under the assumption that disruptions are fixed in at most an hour, which is usually not the case. 

\begin{figure}[h!]
  \centering
  \includegraphics[width=\columnwidth]{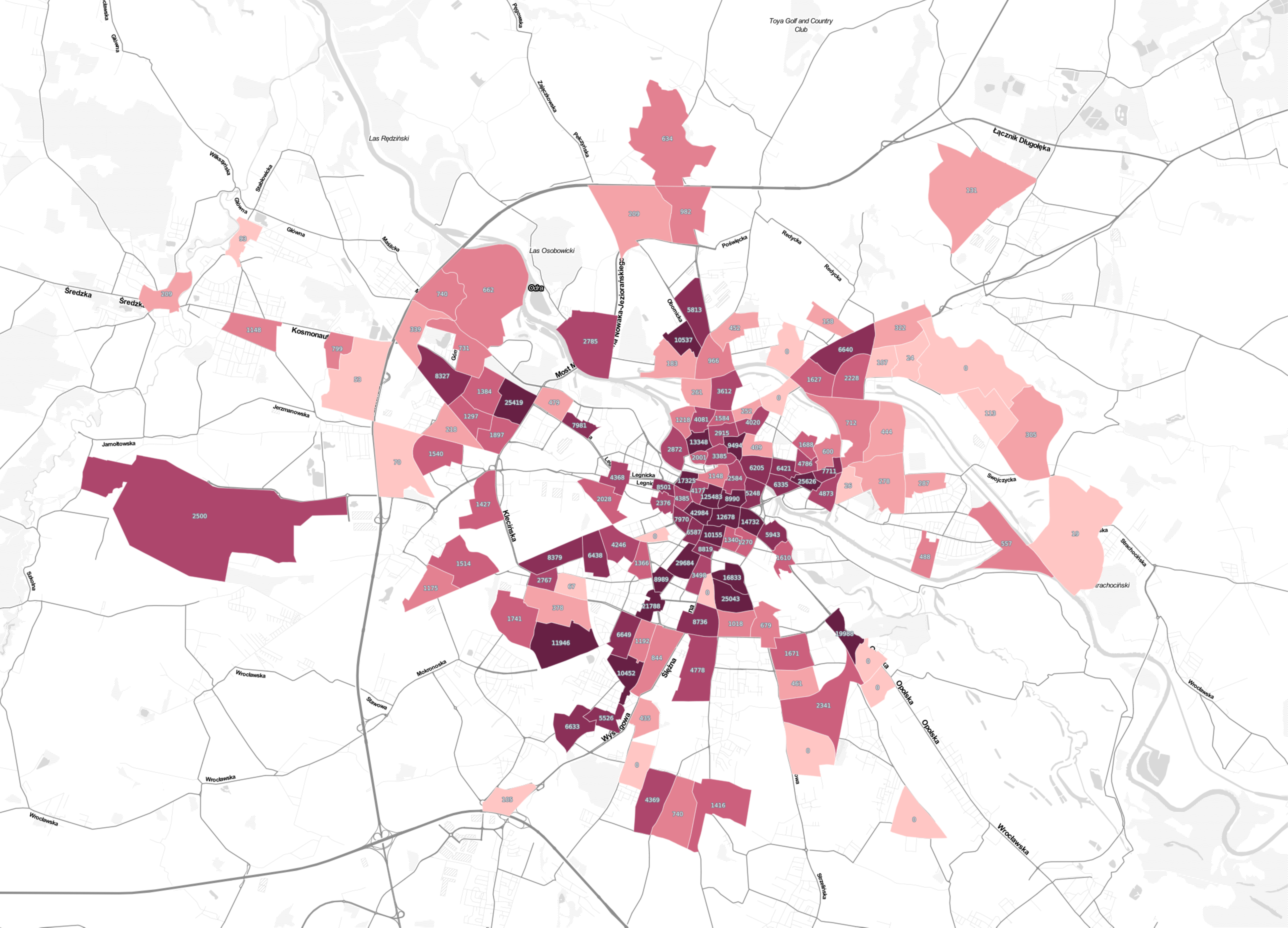}
  \caption{Map of total passengers impacted per region during service disruption on weekdays in the evaluated data set.}
  \label{fig:kbr-passengers-impacted}
\end{figure}

\subsubsection{Sentiment of reactions}

Another use case of our approach is to understand how citizens and passengers react to the incidents. The sentiment of the text data differs when comparing MPK posts and users' comments referring to the posts. The Facebook comments section, although moderated in the case of MPK profile, consists of texts of multiple styles, tones, lengths, which are expressing different attitudes towards the incidents and events described in the posts as well as towards the public transport company (MPK). Thus we found it interesting to evaluate the sentiment of the comments.

Detecting sentiment of passengers proved to be an arduous task due to the presence of misspelling, grammatically incorrect sentences, and contextualized writing style, that assume possessing the appropriate knowledge to interpret the content. Regular incidents caused completely different reactions among passengers, starting from specifying the details of the event and ending with sarcasm and humor. There is a convention that passengers use positive words to carry a negative message. It varies from person to person and depends on many aspects such as age, gender and others. 

The difficulty of the task may be evidenced by the fact, that during the process of manual labeling, the annotators were relatively often conflicted about the right class label for a comment, which was measured by the kappa score between the pairs of annotators (the value ranged from 0.3 to 0.45 and was much lower than the same measure for the post type classification - on average 0.77). What made it even more difficult, besides sarcasm and ambiguity, was the length of the comments. Most of them are relatively short. Hence, in the preprocessing, comments containing no more than three words were removed. Comments which consisted only of emojis or pictures were also filtered.

As it is very hard to identify sarcasm and humor in the content. These kinds of human communication are strongly dependent on context and comment context plays an important role in sarcasm and humor polarity detection. For example, comments: \textit{“MPK did not disappoint my expectations”} and \textit{“The new week is starting nicely”} could be perceived in two ways depending on the context: sarcastic and humorous or ordinary. If the topic relates to an accident - negatively, and if it relates to innovation or infrastructure improvement - positively. Finally, the root of sarcasm lies in frustration \cite{sarcasm2016Filik}. Thus there is a very high probability that a sarcastic comment has negative polarity, which is the opposite of humorous content. 

\begin{figure}[h]
  \centering
  \includegraphics[width=\columnwidth]{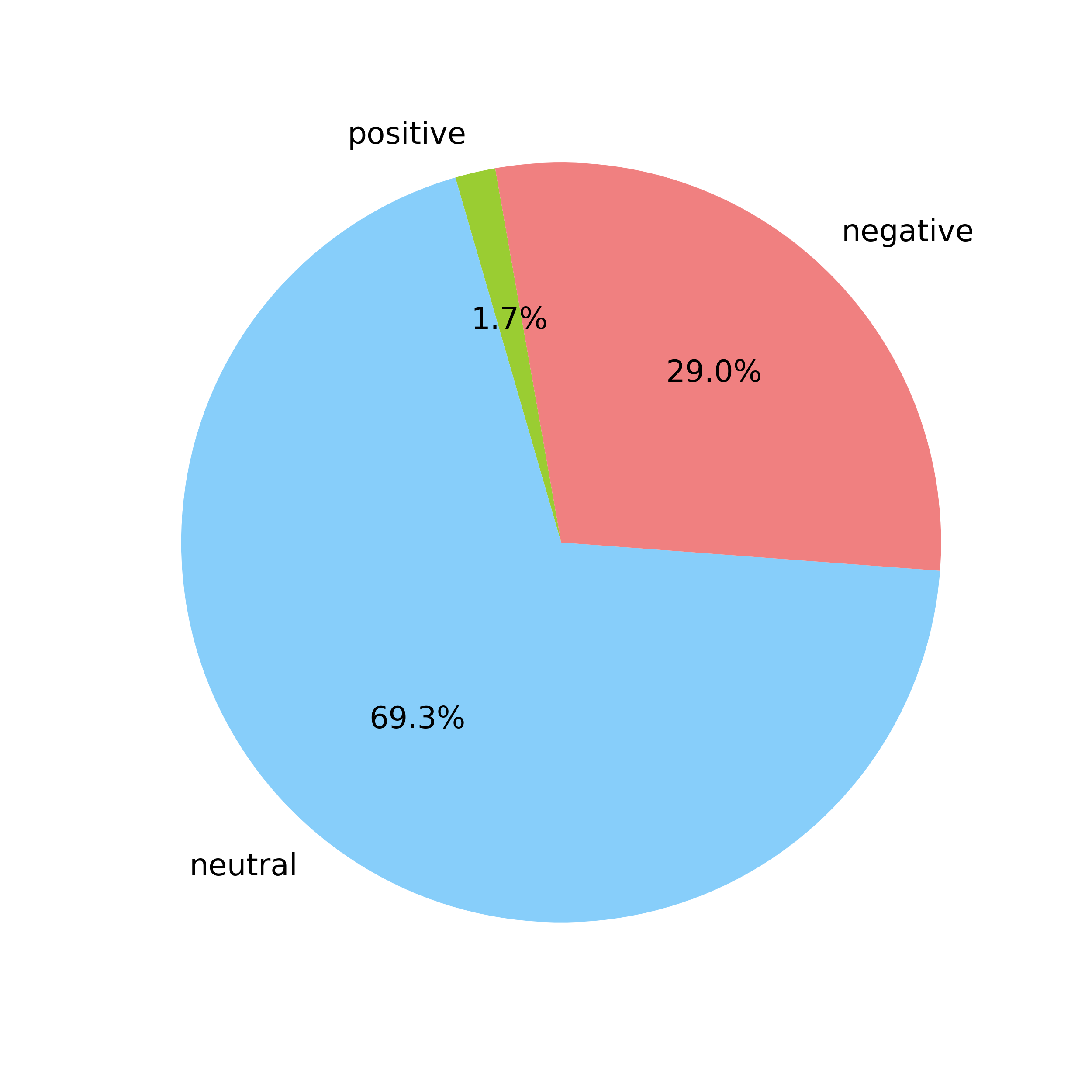}
  \caption{The general sentiment of the comments. The comments were labeled automatically by the fast-text model trained on the set of manually tagged observations.}
  \label{fig:sentiment-pie-chart}
\end{figure}

The basic sentiment analysis classifying comments whether they are positive, neutral or negative resulted in a very imbalanced data set. As figure \ref{fig:sentiment-pie-chart} shows, most of the comments were labeled as neutral. Neutral comments are the comments that are emotionally neutral and the ones that are too short or too ambiguous to assign them as positive or negative. Because of that and the fact that most of the comments are relatively short and thus hard to classify, the neutral class is the most numerous. The least numerous category was a positive class. Since the main topic of the analysis was the comments under the posts about problems in motion, positive comments should not appear often.

\begin{figure}[h]
  \centering
  \includegraphics[width=\columnwidth]{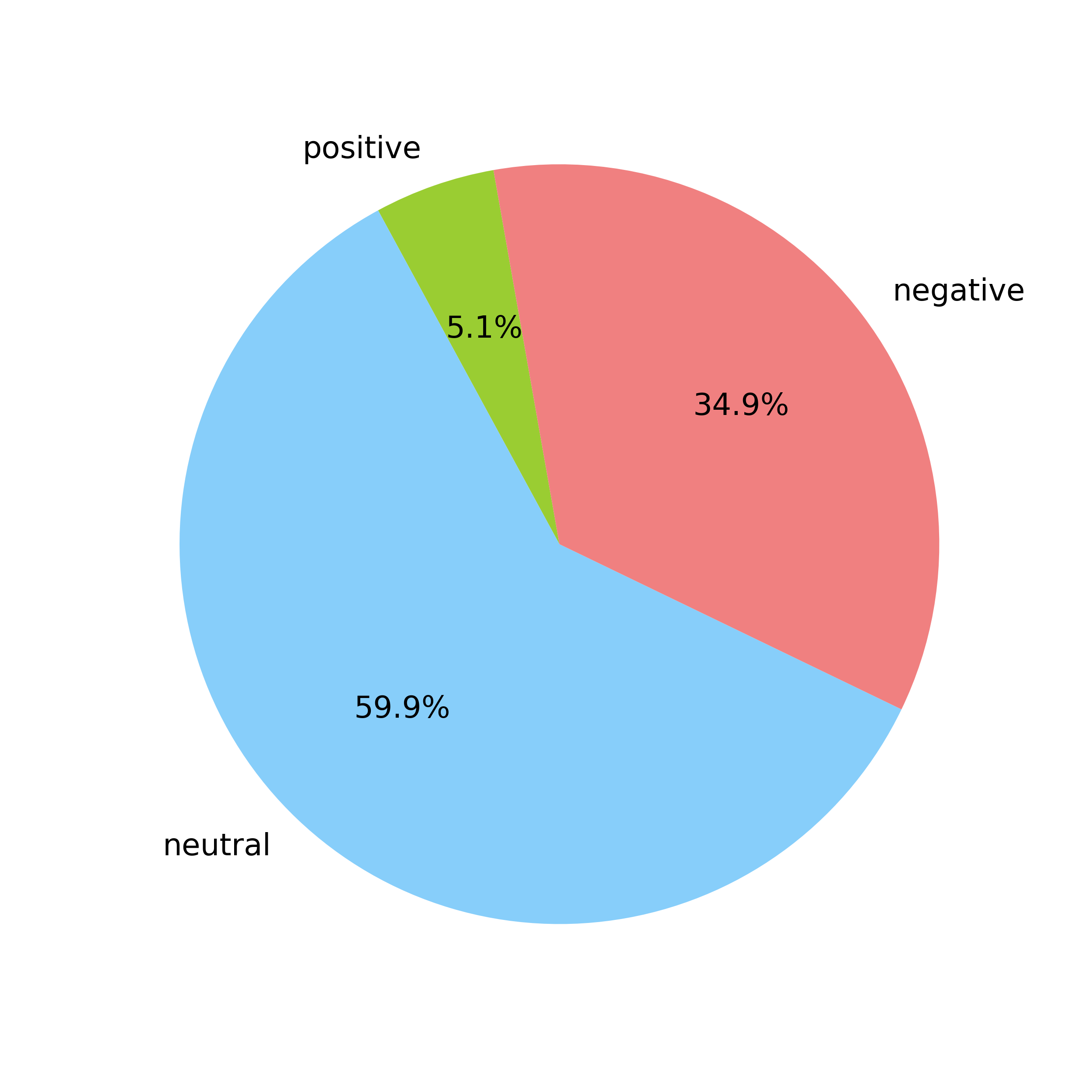}
  \caption{The polarity of the comments towards MPK, where the comments were manually labeled.}
  \label{fig:sentiment-towards-mpk-pie-chart}
\end{figure}

When another approach was tested, which measured the sentiment towards MPK rather than the general emotional attitude expressed in the commentary, positive reactions occurred much more often \ref{fig:sentiment-towards-mpk-pie-chart}. Unfortunately, the percent of negative comments was also higher.

\begin{figure}[h]
  \centering
  \begin{subfigure}{\columnwidth}
    \includegraphics[width=\linewidth]{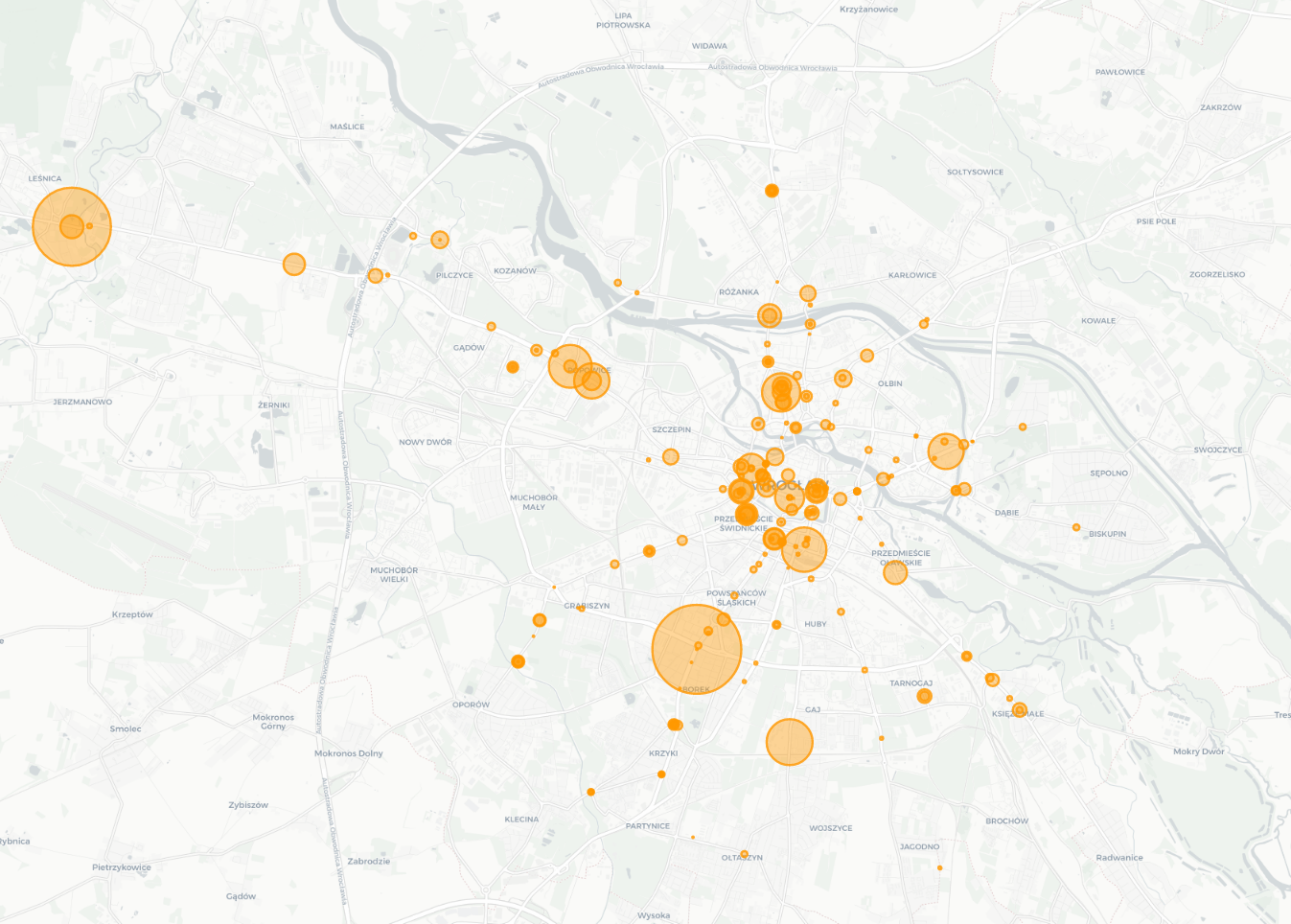}
    \caption{Malfunctions} \label{fig:maps-reactions:a}
  \end{subfigure}
\end{figure}
\begin{figure}[h]\ContinuedFloat
  \centering
  \begin{subfigure}{\columnwidth}
    \includegraphics[width=\linewidth]{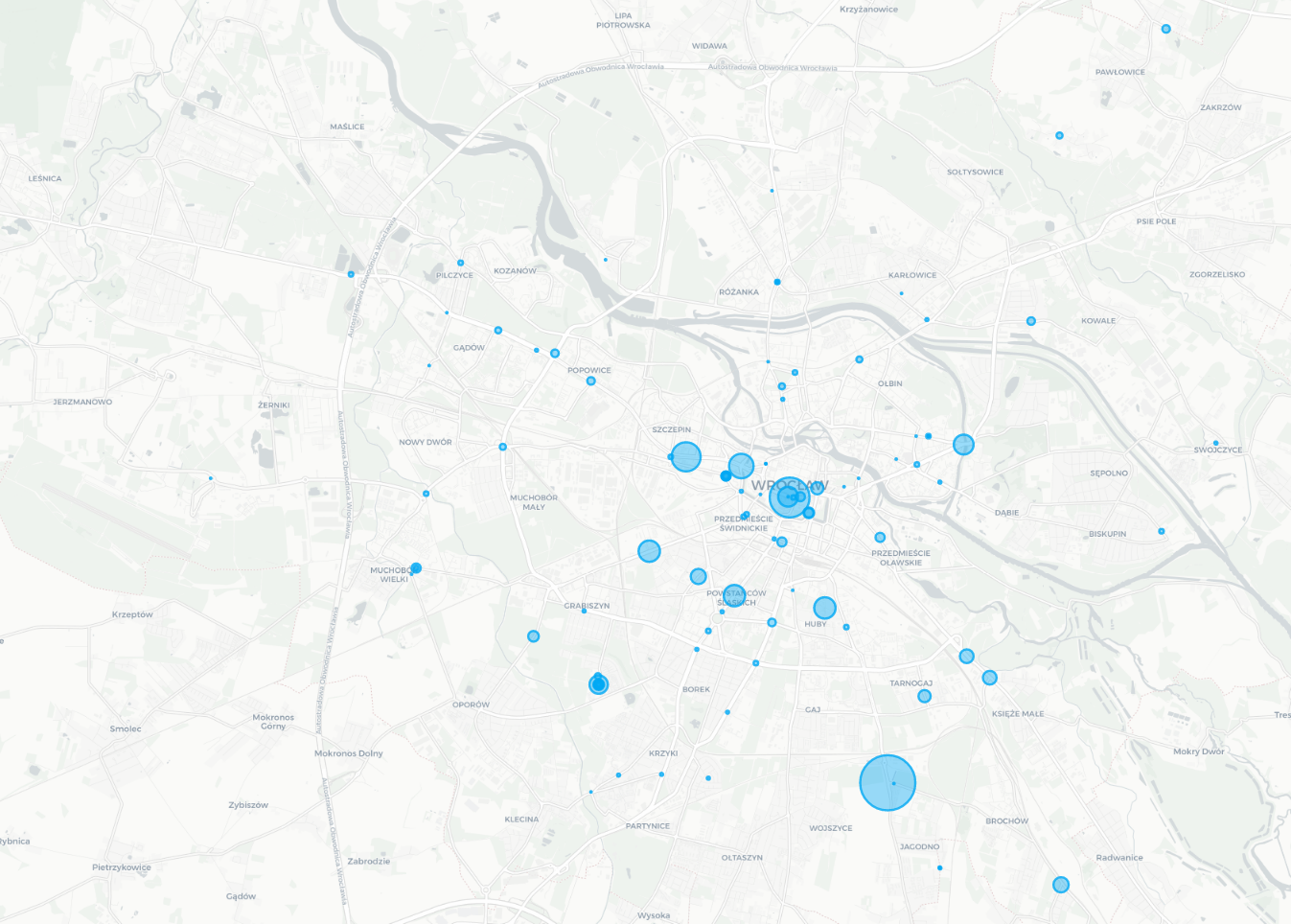}
    \caption{Random incidents} \label{fig:maps-reactions:b}
  \end{subfigure}
\end{figure}
\begin{figure}[h]\ContinuedFloat
  \centering
  \begin{subfigure}{\columnwidth}
    \includegraphics[width=\linewidth]{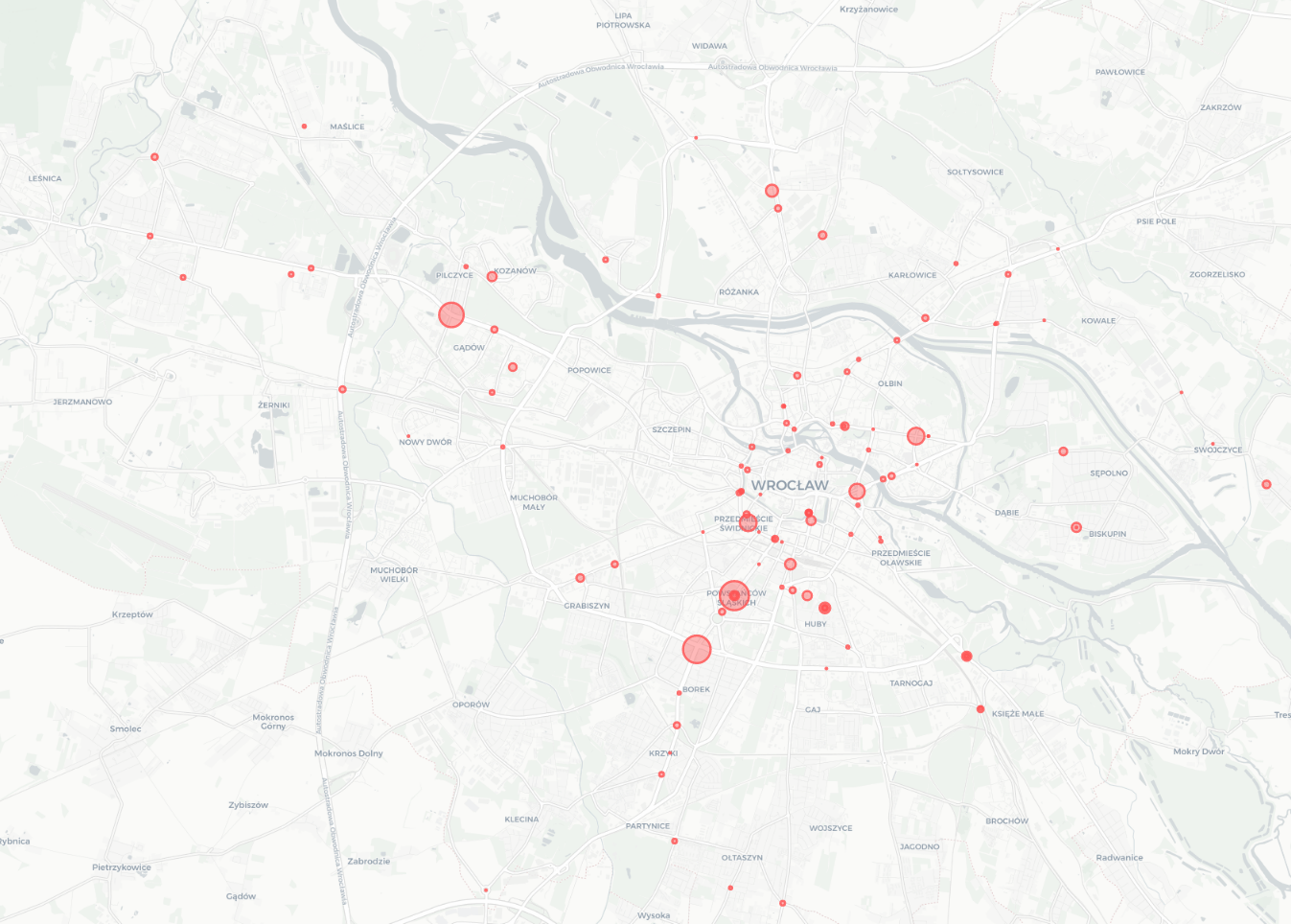}
    \caption{Traffic accidents} \label{fig:maps-reactions:c}
  \end{subfigure}
  \caption{Maps showing disruptions positions and reactions bound to them as circles. From the top: \textbf{malfunctions}, \textbf{incidents}, \textbf{accidents}. \textbf{Unknown} type was left out, because there was almost zero interactions under those posts.}
  \label{fig:maps-reactions}
\end{figure}

\subsubsection{Reactions and spread}

\begin{figure}[h]
    \centering
    \includegraphics[width=\columnwidth]{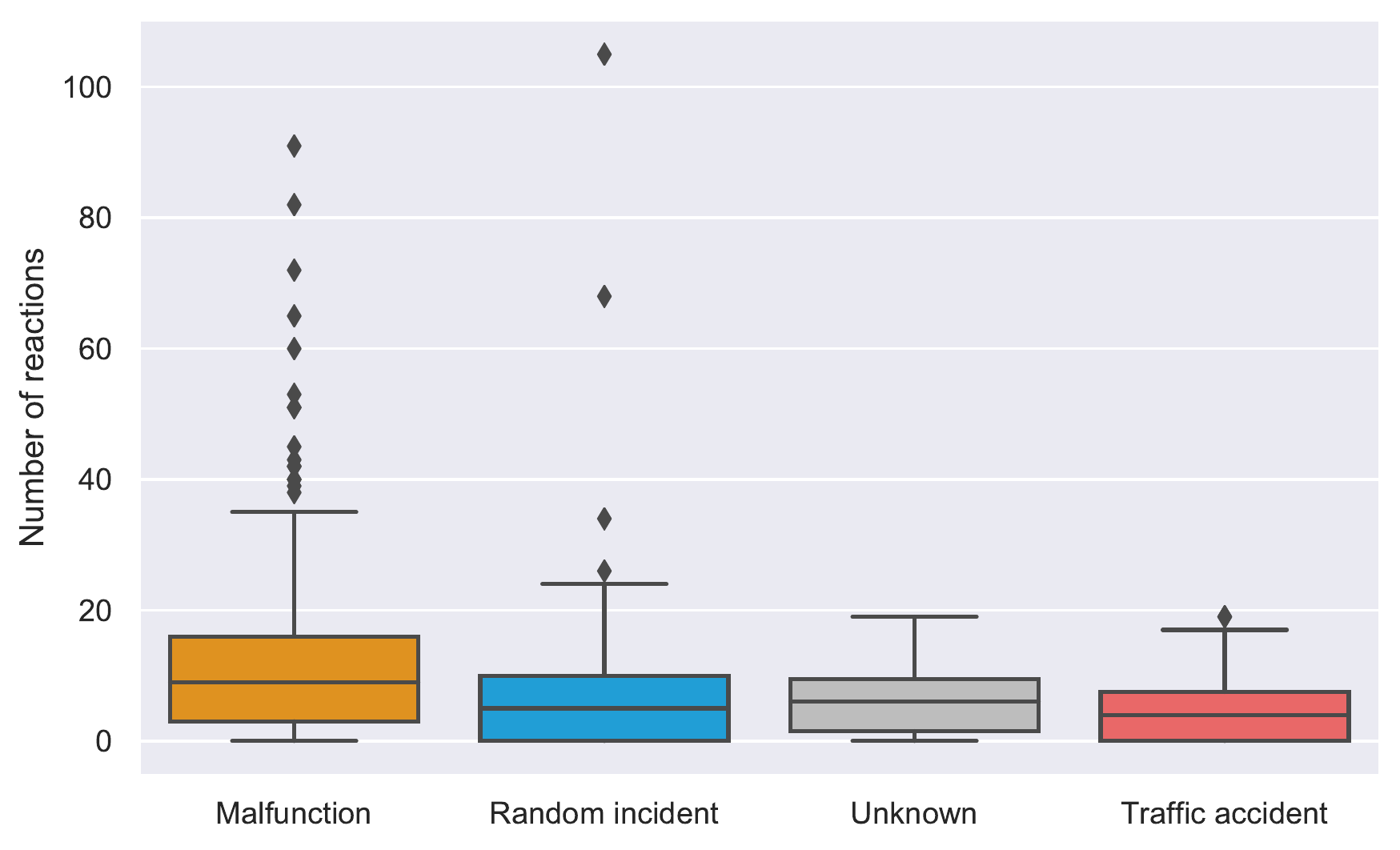}
    \caption[short]{Distribution of reactions to \#AlertMPK posts per type.}
    \label{fig:reactions-chart}
\end{figure}

User's reactions can help approximate social sentiment towards certain types of disruptions.
First of all, "Like" reaction is most apparent in the list probably because it's the easiest and most popular reaction for Facebook users. Additionally, there is a significant difference between "Haha" reaction and the rest of the reactions. Figure \ref{fig:reactions-chart} shows full comparison of number of reactions to \#AlertMPK posts. 

Maps in figure \ref{fig:maps-reactions} show disruption positions marked as circles where the radius of a circle is bound to a total amount of passengers' interactions (reactions + number of comments) with a post. The visualization shows that people react most to malfunctions and least to traffic accidents. Few outliers with significantly more reactions were found, especially among \textit{malfunction} type of posts. We describe in more detail three posts with the highest number of interactions.

\begin{figure}[h]
\centering
\includegraphics[width=\columnwidth]{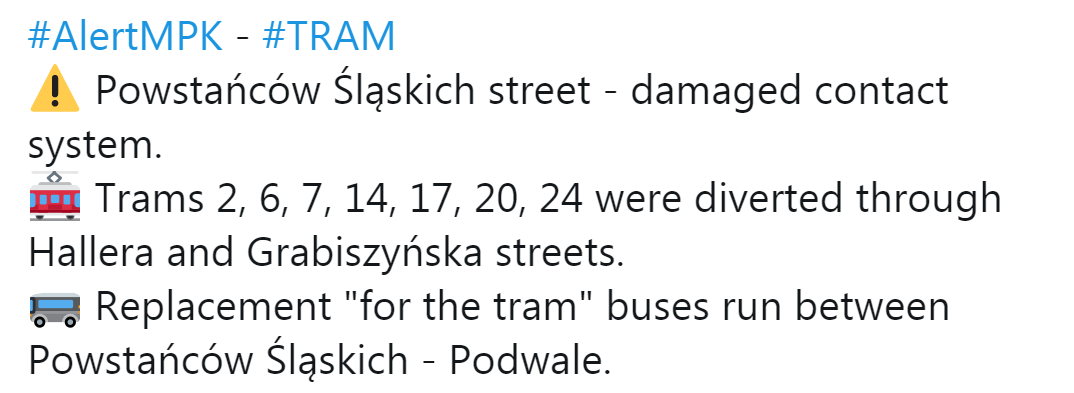}
\caption[short]{Post of type \textbf{Malfunction} with 82 reactions and 144 comments (226 interactions total). Can be seen in the south of the map \ref{fig:maps-reactions:a}.} \label{fig:top-reacted-posts:a}
\end{figure}

First disruption (Figure \ref{fig:top-reacted-posts:a}) blocked south part of a city on tram rails where 7 tram lines run every day. Diverted trams increased delay times of other tram lines and replacement buses hugely impacted traffic jams because Powstańców Śląskich street is one of the busiest streets in the city. Cause of disruption was a broken tram pantograph and people were mad, that's why there are more comments than reactions under this post.

\begin{figure}[h]
\centering
\includegraphics[width=\columnwidth]{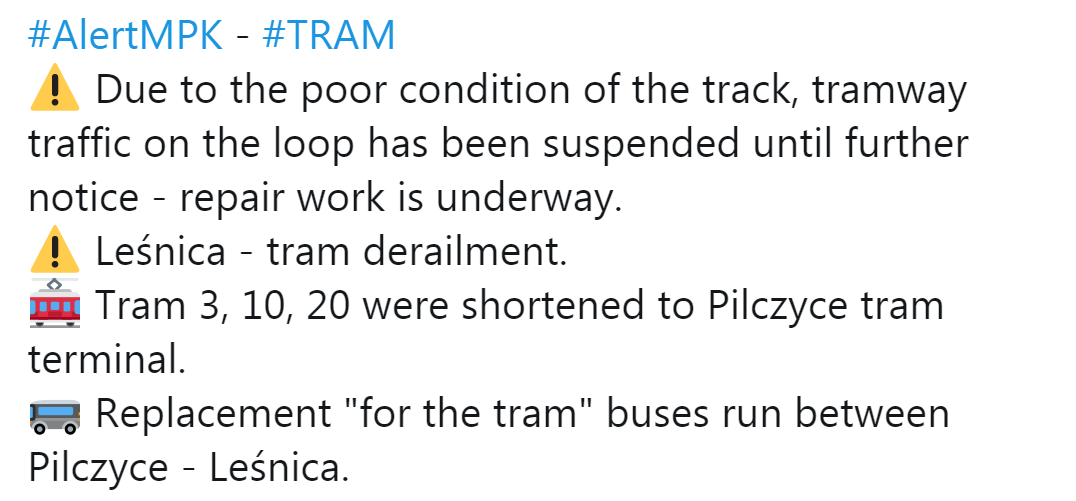}
\caption[short]{Post of type \textbf{Malfunction} with 91 reactions and 109 comments (200 interactions total). Can be seen in the north-west of the map \ref{fig:maps-reactions:a}.} \label{fig:top-reacted-posts:b}
\end{figure}

Second disruption (Figure \ref{fig:top-reacted-posts:b}) disabled the only tram track leading to one of the furthest districts in the city - Leśnica. Replacement buses had to keep waiting in a long traffic jam and many people who depended on a public transport system were even hours later at home after work. Same as first disruption, the cause of it was a malfunction and again passengers' frustration was reflected in a high number of comments demanding to improve the condition of public transport.

\begin{figure}[h]
\centering
\includegraphics[width=\columnwidth]{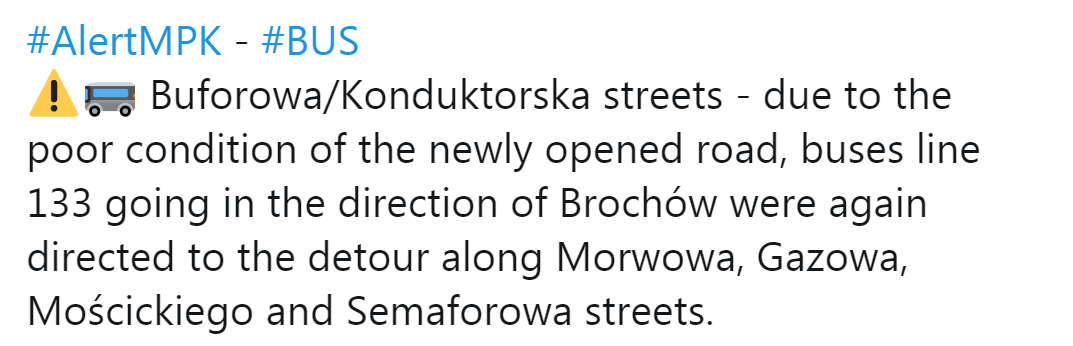}
\caption[short]{Post of type \textbf{Incident} with 105 reactions and 35 comments (140 interactions total). Can be seen in the south-east of the map \ref{fig:maps-reactions:b}.} \label{fig:top-reacted-posts:c}
\end{figure}

The third disruption (Figure \ref{fig:top-reacted-posts:c}) was independent of MPK and was depended on poorly performed road works. That's why there is a big difference between the number of reactions (105) and comments (35). Additionally, most of these reactions were "Haha" reactions (77). The disruption amused passengers more than upset them.

\section{Conclusions}
To address the problem of spatial understanding of social media posts that convey transport information, we collected a three-year data set of public posts from official channels of a polish public transport company. After exploration, we proposed a systematic typology of communique and discussed the magnitude of social response, which may be useful for building similar tools for other agencies. For this typology, we build a successful classifier using language models at the time available for the Polish language. We used lists of street names and stop names from open data sources such as GTFS and OpenStreetMap. We used distance-based matching of word sequences in a manually selected context to provide spatial understanding. We then managed to successfully use the obtained spatial understanding to relate information from social posts to external data sets such as the mobility survey. We applied our approach to discussing the aggregated impact on the transportation network and mobility and automatic examination of the sentiment of citizens' feedback. We release both the data sets and experimental codebase under a permissive opensource license to encourage further research in the topic, especially in under-resourced languages and to empower data scientists worldwide with a tool for performing citizen science.

\bibliographystyle{ACM-Reference-Format}
\bibliography{sample-base}


\begin{thebibliography}{25}


\ifx \showCODEN    \undefined \def \showCODEN     #1{\unskip}     \fi
\ifx \showDOI      \undefined \def \showDOI       #1{#1}\fi
\ifx \showISBNx    \undefined \def \showISBNx     #1{\unskip}     \fi
\ifx \showISBNxiii \undefined \def \showISBNxiii  #1{\unskip}     \fi
\ifx \showISSN     \undefined \def \showISSN      #1{\unskip}     \fi
\ifx \showLCCN     \undefined \def \showLCCN      #1{\unskip}     \fi
\ifx \shownote     \undefined \def \shownote      #1{#1}          \fi
\ifx \showarticletitle \undefined \def \showarticletitle #1{#1}   \fi
\ifx \showURL      \undefined \def \showURL       {\relax}        \fi
\providecommand\bibfield[2]{#2}
\providecommand\bibinfo[2]{#2}
\providecommand\natexlab[1]{#1}
\providecommand\showeprint[2][]{arXiv:#2}

\bibitem[\protect\citeauthoryear{Alshehri and O’Keefe}{Alshehri and
  O’Keefe}{2019}]%
        {alshehri2019analyzing}
\bibfield{author}{\bibinfo{person}{Alanawd Alshehri} {and}
  \bibinfo{person}{Robert O’Keefe}.} \bibinfo{year}{2019}\natexlab{}.
\newblock \showarticletitle{Analyzing Social Media to Assess User Satisfaction
  with Transport for London’s Oyster}.
\newblock \bibinfo{journal}{\emph{International Journal of Human--Computer
  Interaction}} \bibinfo{volume}{35}, \bibinfo{number}{15}
  (\bibinfo{year}{2019}), \bibinfo{pages}{1378--1387}.
\newblock


\bibitem[\protect\citeauthoryear{Bregman}{Bregman}{2012}]%
        {bregman2012uses}
\bibfield{author}{\bibinfo{person}{Susan Bregman}.}
  \bibinfo{year}{2012}\natexlab{}.
\newblock \bibinfo{booktitle}{\emph{Uses of social media in public
  transportation}}.
\newblock Number~99. \bibinfo{publisher}{Transportation Research Board}.
\newblock


\bibitem[\protect\citeauthoryear{Chaniotakis, Antoniou, Aifadopoulou, and
  Dimitriou}{Chaniotakis et~al\mbox{.}}{2017}]%
        {chaniotakis2017inferring}
\bibfield{author}{\bibinfo{person}{Emmanouil Chaniotakis},
  \bibinfo{person}{Constantinos Antoniou}, \bibinfo{person}{Georgia
  Aifadopoulou}, {and} \bibinfo{person}{Loukas Dimitriou}.}
  \bibinfo{year}{2017}\natexlab{}.
\newblock \showarticletitle{Inferring activities from social media data}.
\newblock \bibinfo{journal}{\emph{Transportation research record}}
  \bibinfo{volume}{2666}, \bibinfo{number}{1} (\bibinfo{year}{2017}),
  \bibinfo{pages}{29--37}.
\newblock


\bibitem[\protect\citeauthoryear{Congosto, Fuentes-Lorenzo, and
  S{\'a}nchez}{Congosto et~al\mbox{.}}{2015}]%
        {congosto2015microbloggers}
\bibfield{author}{\bibinfo{person}{Mariluz Congosto}, \bibinfo{person}{Damaris
  Fuentes-Lorenzo}, {and} \bibinfo{person}{Luis S{\'a}nchez}.}
  \bibinfo{year}{2015}\natexlab{}.
\newblock \showarticletitle{Microbloggers as sensors for public transport
  breakdowns}.
\newblock \bibinfo{journal}{\emph{IEEE Internet Computing}}
  \bibinfo{volume}{19}, \bibinfo{number}{6} (\bibinfo{year}{2015}),
  \bibinfo{pages}{18--25}.
\newblock


\bibitem[\protect\citeauthoryear{Cottrill, Gault, Yeboah, Nelson, Anable, and
  Budd}{Cottrill et~al\mbox{.}}{2017}]%
        {cottrill2017tweeting}
\bibfield{author}{\bibinfo{person}{Caitlin Cottrill}, \bibinfo{person}{Paul
  Gault}, \bibinfo{person}{Godwin Yeboah}, \bibinfo{person}{John~D Nelson},
  \bibinfo{person}{Jillian Anable}, {and} \bibinfo{person}{Thomas Budd}.}
  \bibinfo{year}{2017}\natexlab{}.
\newblock \showarticletitle{Tweeting Transit: An examination of social media
  strategies for transport information management during a large event}.
\newblock \bibinfo{journal}{\emph{Transportation Research Part C: Emerging
  Technologies}}  \bibinfo{volume}{77} (\bibinfo{year}{2017}),
  \bibinfo{pages}{421--432}.
\newblock


\bibitem[\protect\citeauthoryear{Filik, Turcan, Thompson, Harvey, Davies, and
  Turner}{Filik et~al\mbox{.}}{2016}]%
        {sarcasm2016Filik}
\bibfield{author}{\bibinfo{person}{Ruth Filik}, \bibinfo{person}{Alexandra
  Turcan}, \bibinfo{person}{Dominic Thompson}, \bibinfo{person}{Nicole Harvey},
  \bibinfo{person}{Harriet Davies}, {and} \bibinfo{person}{Amelia Turner}.}
  \bibinfo{year}{2016}\natexlab{}.
\newblock \showarticletitle{Sarcasm and emoticons: Comprehension and emotional
  impact}.
\newblock \bibinfo{journal}{\emph{Quarterly Journal of Experimental
  Psychology}}  \bibinfo{volume}{69} (\bibinfo{year}{2016}),
  \bibinfo{pages}{2130 -- 2146}.
\newblock
\urldef\tempurl%
\url{https://doi.org/10.1080/17470218.2015.1106566}
\showDOI{\tempurl}


\bibitem[\protect\citeauthoryear{Gal-Tzur, Grant-Muller, Kuflik, Minkov,
  Nocera, and Shoor}{Gal-Tzur et~al\mbox{.}}{2014}]%
        {GALTZUR2014115}
\bibfield{author}{\bibinfo{person}{Ayelet Gal-Tzur}, \bibinfo{person}{Susan~M.
  Grant-Muller}, \bibinfo{person}{Tsvi Kuflik}, \bibinfo{person}{Einat Minkov},
  \bibinfo{person}{Silvio Nocera}, {and} \bibinfo{person}{Itay Shoor}.}
  \bibinfo{year}{2014}\natexlab{}.
\newblock \showarticletitle{The potential of social media in delivering
  transport policy goals}.
\newblock \bibinfo{journal}{\emph{Transport Policy}}  \bibinfo{volume}{32}
  (\bibinfo{year}{2014}), \bibinfo{pages}{115 -- 123}.
\newblock
\showISSN{0967-070X}
\urldef\tempurl%
\url{https://doi.org/10.1016/j.tranpol.2014.01.007}
\showDOI{\tempurl}


\bibitem[\protect\citeauthoryear{Gault, Cottrill, Corsar, Edwards, Nelson,
  Markovic, Mehdi, and Sripada}{Gault et~al\mbox{.}}{2019}]%
        {gault2019travelbot}
\bibfield{author}{\bibinfo{person}{Paul Gault}, \bibinfo{person}{Caitlin~D
  Cottrill}, \bibinfo{person}{David Corsar}, \bibinfo{person}{Peter Edwards},
  \bibinfo{person}{John~D Nelson}, \bibinfo{person}{Milan Markovic},
  \bibinfo{person}{Mujtaba Mehdi}, {and} \bibinfo{person}{Somayajulu Sripada}.}
  \bibinfo{year}{2019}\natexlab{}.
\newblock \showarticletitle{TravelBot: Utilising social media dialogue to
  provide journey disruption alerts}.
\newblock \bibinfo{journal}{\emph{Transportation Research Interdisciplinary
  Perspectives}}  \bibinfo{volume}{3} (\bibinfo{year}{2019}),
  \bibinfo{pages}{100062}.
\newblock


\bibitem[\protect\citeauthoryear{Haghighi, Liu, Wei, Li, and Shao}{Haghighi
  et~al\mbox{.}}{2018}]%
        {haghighi2018using}
\bibfield{author}{\bibinfo{person}{N~Nima Haghighi},
  \bibinfo{person}{Xiaoyue~Cathy Liu}, \bibinfo{person}{Ran Wei},
  \bibinfo{person}{Wenwen Li}, {and} \bibinfo{person}{Hu Shao}.}
  \bibinfo{year}{2018}\natexlab{}.
\newblock \showarticletitle{Using Twitter data for transit performance
  assessment: a framework for evaluating transit riders’ opinions about
  quality of service}.
\newblock \bibinfo{journal}{\emph{Public Transport}} \bibinfo{volume}{10},
  \bibinfo{number}{2} (\bibinfo{year}{2018}), \bibinfo{pages}{363--377}.
\newblock


\bibitem[\protect\citeauthoryear{Joulin, Grave, Bojanowski, and Mikolov}{Joulin
  et~al\mbox{.}}{2017}]%
        {joulin2017bag}
\bibfield{author}{\bibinfo{person}{Armand Joulin}, \bibinfo{person}{Edouard
  Grave}, \bibinfo{person}{Piotr Bojanowski}, {and} \bibinfo{person}{Tomas
  Mikolov}.} \bibinfo{year}{2017}\natexlab{}.
\newblock \showarticletitle{Bag of Tricks for Efficient Text Classification}.
  In \bibinfo{booktitle}{\emph{Proceedings of the 15th Conference of the
  European Chapter of the Association for Computational Linguistics: Volume 2,
  Short Papers}}. \bibinfo{publisher}{Association for Computational
  Linguistics}, \bibinfo{pages}{427--431}.
\newblock


\bibitem[\protect\citeauthoryear{Li, Zhang, and Li}{Li et~al\mbox{.}}{2019}]%
        {li2019mining}
\bibfield{author}{\bibinfo{person}{Dawei Li}, \bibinfo{person}{Yujia Zhang},
  {and} \bibinfo{person}{Cheng Li}.} \bibinfo{year}{2019}\natexlab{}.
\newblock \showarticletitle{Mining Public Opinion on Transportation Systems
  Based on Social Media Data}.
\newblock \bibinfo{journal}{\emph{Sustainability}} \bibinfo{volume}{11},
  \bibinfo{number}{15} (\bibinfo{year}{2019}), \bibinfo{pages}{1--15}.
\newblock
\showISSN{20711050}
\urldef\tempurl%
\url{https://doi.org/10.3390/su11154016}
\showDOI{\tempurl}


\bibitem[\protect\citeauthoryear{Liu, Shi, Elrahman, Xuegang, and Reilly}{Liu
  et~al\mbox{.}}{2016}]%
        {lui2016understanding}
\bibfield{author}{\bibinfo{person}{Jenny Liu}, \bibinfo{person}{Wei Shi},
  \bibinfo{person}{Sam Elrahman}, \bibinfo{person}{Ban Xuegang}, {and}
  \bibinfo{person}{Jack Reilly}.} \bibinfo{year}{2016}\natexlab{}.
\newblock \showarticletitle{Understanding social media program usage in public
  transit agencies}.
\newblock \bibinfo{journal}{\emph{International Journal of Transportation
  Science and Technology}}  \bibinfo{volume}{2} (\bibinfo{year}{2016}),
  \bibinfo{pages}{83--92}.
\newblock
\showISSN{20460449}
\urldef\tempurl%
\url{https://doi.org/10.1016/j.ijtst.2016.09.005}
\showDOI{\tempurl}


\bibitem[\protect\citeauthoryear{Lv, Chen, Zhang, Duan, and Li}{Lv
  et~al\mbox{.}}{2017}]%
        {lv2017social}
\bibfield{author}{\bibinfo{person}{Yisheng Lv}, \bibinfo{person}{Yuanyuan
  Chen}, \bibinfo{person}{Xiqiao Zhang}, \bibinfo{person}{Yanjie Duan}, {and}
  \bibinfo{person}{Naiqiang~Li Li}.} \bibinfo{year}{2017}\natexlab{}.
\newblock \showarticletitle{Social media based transportation research: The
  state of the work and the networking}.
\newblock \bibinfo{journal}{\emph{IEEE/CAA Journal of Automatica Sinica}}
  \bibinfo{volume}{4}, \bibinfo{number}{1} (\bibinfo{year}{2017}),
  \bibinfo{pages}{19--26}.
\newblock


\bibitem[\protect\citeauthoryear{Mogridge, Holden, Bird, and Terzis}{Mogridge
  et~al\mbox{.}}{1987}]%
        {mogridge1987downs}
\bibfield{author}{\bibinfo{person}{Martin John~Henry Mogridge},
  \bibinfo{person}{DJ Holden}, \bibinfo{person}{J Bird}, {and}
  \bibinfo{person}{GC Terzis}.} \bibinfo{year}{1987}\natexlab{}.
\newblock \showarticletitle{The Downs/Thomson paradox and the transportation
  planning process}.
\newblock \bibinfo{journal}{\emph{International Journal of Transport
  Economics/Rivista internazionale di economia dei trasporti}}
  (\bibinfo{year}{1987}), \bibinfo{pages}{283--311}.
\newblock


\bibitem[\protect\citeauthoryear{Ni, He, and Gao}{Ni et~al\mbox{.}}{2016}]%
        {ni2016forecasting}
\bibfield{author}{\bibinfo{person}{Ming Ni}, \bibinfo{person}{Qing He}, {and}
  \bibinfo{person}{Jing Gao}.} \bibinfo{year}{2016}\natexlab{}.
\newblock \showarticletitle{Forecasting the subway passenger flow under event
  occurrences with social media}.
\newblock \bibinfo{journal}{\emph{IEEE Transactions on Intelligent
  Transportation Systems}} \bibinfo{volume}{18}, \bibinfo{number}{6}
  (\bibinfo{year}{2016}), \bibinfo{pages}{1623--1632}.
\newblock


\bibitem[\protect\citeauthoryear{Nikolaidou and Papaioannou}{Nikolaidou and
  Papaioannou}{2018}]%
        {nikolaidou2018utilizing}
\bibfield{author}{\bibinfo{person}{Anastasia Nikolaidou} {and}
  \bibinfo{person}{Panagiotis Papaioannou}.} \bibinfo{year}{2018}\natexlab{}.
\newblock \showarticletitle{Utilizing social media in transport planning and
  public transit quality: Survey of literature}.
\newblock \bibinfo{journal}{\emph{Journal of Transportation Engineering, Part
  A: Systems}} \bibinfo{volume}{144}, \bibinfo{number}{4}
  (\bibinfo{year}{2018}), \bibinfo{pages}{04018007}.
\newblock


\bibitem[\protect\citeauthoryear{Pender, Currie, Delbosc, and Shiwakoti}{Pender
  et~al\mbox{.}}{2014a}]%
        {doi:10.1080/01441647.2014.915442}
\bibfield{author}{\bibinfo{person}{Brendan Pender}, \bibinfo{person}{Graham
  Currie}, \bibinfo{person}{Alexa Delbosc}, {and} \bibinfo{person}{Nirajan
  Shiwakoti}.} \bibinfo{year}{2014}\natexlab{a}.
\newblock \showarticletitle{Social Media Use during Unplanned Transit Network
  Disruptions: A Review of Literature}.
\newblock \bibinfo{journal}{\emph{Transport Reviews}} \bibinfo{volume}{34},
  \bibinfo{number}{4} (\bibinfo{year}{2014}), \bibinfo{pages}{501--521}.
\newblock
\urldef\tempurl%
\url{https://doi.org/10.1080/01441647.2014.915442}
\showDOI{\tempurl}


\bibitem[\protect\citeauthoryear{Pender, Currie, Delbosc, and Shiwakoti}{Pender
  et~al\mbox{.}}{2014b}]%
        {pender2014social}
\bibfield{author}{\bibinfo{person}{Brendan Pender}, \bibinfo{person}{Graham
  Currie}, \bibinfo{person}{Alexa Delbosc}, {and} \bibinfo{person}{Nirajan
  Shiwakoti}.} \bibinfo{year}{2014}\natexlab{b}.
\newblock \showarticletitle{Social media use during unplanned transit network
  disruptions: A review of literature}.
\newblock \bibinfo{journal}{\emph{Transport Reviews}} \bibinfo{volume}{34},
  \bibinfo{number}{4} (\bibinfo{year}{2014}), \bibinfo{pages}{501--521}.
\newblock


\bibitem[\protect\citeauthoryear{Perriam}{Perriam}{2019}]%
        {perriam2019}
\bibfield{author}{\bibinfo{person}{Jessamy Perriam}.}
  \bibinfo{year}{2019}\natexlab{}.
\newblock \showarticletitle{A Tweet is Not Just a Tweet: Public Sector
  Understandings and Analysis of Social Media Customer Service Data}.
\newblock \bibinfo{journal}{\emph{SMSociety '19: Proceedings of the 10th
  International Conference on Social Media and Society}},
  \bibinfo{pages}{33--40}.
\newblock
\showISBNx{978-1-4503-6651-9}
\urldef\tempurl%
\url{https://doi.org/10.1145/3328529.3328542}
\showDOI{\tempurl}


\bibitem[\protect\citeauthoryear{Raghothama, Magal~Shreenath, and
  Meijer}{Raghothama et~al\mbox{.}}{2016}]%
        {ragothama2016analytics}
\bibfield{author}{\bibinfo{person}{Jayanth Raghothama},
  \bibinfo{person}{Vinutha Magal~Shreenath}, {and} \bibinfo{person}{Sebastiaan
  Meijer}.} \bibinfo{year}{2016}\natexlab{}.
\newblock \showarticletitle{Analytics on public transport delays with spatial
  big data}. \bibinfo{pages}{28--33}.
\newblock
\urldef\tempurl%
\url{https://doi.org/10.1145/3006386.3006387}
\showDOI{\tempurl}


\bibitem[\protect\citeauthoryear{Rashidi, Abbasi, Maghrebi, Hasan, and
  Waller}{Rashidi et~al\mbox{.}}{2017}]%
        {rashidi2017exploring}
\bibfield{author}{\bibinfo{person}{Taha~H Rashidi}, \bibinfo{person}{Alireza
  Abbasi}, \bibinfo{person}{Mojtaba Maghrebi}, \bibinfo{person}{Samiul Hasan},
  {and} \bibinfo{person}{Travis~S Waller}.} \bibinfo{year}{2017}\natexlab{}.
\newblock \showarticletitle{Exploring the capacity of social media data for
  modelling travel behaviour: Opportunities and challenges}.
\newblock \bibinfo{journal}{\emph{Transportation Research Part C: Emerging
  Technologies}}  \bibinfo{volume}{75} (\bibinfo{year}{2017}),
  \bibinfo{pages}{197--211}.
\newblock


\bibitem[\protect\citeauthoryear{Serna-Nocedal, Gerrikagoitia, Bernab{\'e}, and
  Ruiz~S{\'a}nchez}{Serna-Nocedal et~al\mbox{.}}{2017}]%
        {serna2017sustainability}
\bibfield{author}{\bibinfo{person}{Ainhoa Serna-Nocedal},
  \bibinfo{person}{Jon~Kepa Gerrikagoitia}, \bibinfo{person}{Unai Bernab{\'e}},
  {and} \bibinfo{person}{Tom{\'a}s Ruiz~S{\'a}nchez}.}
  \bibinfo{year}{2017}\natexlab{}.
\newblock \showarticletitle{Sustainability analysis on Urban Mobility based on
  Social Media content}.
\newblock \bibinfo{journal}{\emph{Transportation Research Procedia}}
  \bibinfo{volume}{24} (\bibinfo{year}{2017}), \bibinfo{pages}{1--8}.
\newblock


\bibitem[\protect\citeauthoryear{Tu, Cao, Yue, Shaw, Zhou, Wang, Chang, Xu, and
  Li}{Tu et~al\mbox{.}}{2017}]%
        {tu2017coupling}
\bibfield{author}{\bibinfo{person}{Wei Tu}, \bibinfo{person}{Jinzhou Cao},
  \bibinfo{person}{Yang Yue}, \bibinfo{person}{Shih-Lung Shaw},
  \bibinfo{person}{Meng Zhou}, \bibinfo{person}{Zhensheng Wang},
  \bibinfo{person}{Xiaomeng Chang}, \bibinfo{person}{Yang Xu}, {and}
  \bibinfo{person}{Qingquan Li}.} \bibinfo{year}{2017}\natexlab{}.
\newblock \showarticletitle{Coupling mobile phone and social media data: A new
  approach to understanding urban functions and diurnal patterns}.
\newblock \bibinfo{journal}{\emph{International Journal of Geographical
  Information Science}} \bibinfo{volume}{31}, \bibinfo{number}{12}
  (\bibinfo{year}{2017}), \bibinfo{pages}{2331--2358}.
\newblock


\bibitem[\protect\citeauthoryear{Wojtowicz and Wallace}{Wojtowicz and
  Wallace}{2016}]%
        {jeffrey2016management}
\bibfield{author}{\bibinfo{person}{Jeffrey Wojtowicz} {and}
  \bibinfo{person}{William Wallace}.} \bibinfo{year}{2016}\natexlab{}.
\newblock \showarticletitle{Use of social media by transportation agencies for
  traffic management}.
\newblock \bibinfo{journal}{\emph{Transportation Research Record}}
  \bibinfo{volume}{2551} (\bibinfo{year}{2016}), \bibinfo{pages}{82--89}.
\newblock
\showISSN{03611981}
\urldef\tempurl%
\url{https://doi.org/10.3141/2551-10}
\showDOI{\tempurl}


\bibitem[\protect\citeauthoryear{Wrocław}{Wrocław}{2020}]%
        {KBR2018}
\bibfield{author}{\bibinfo{person}{Municipal~Office Wrocław}.}
  \bibinfo{year}{2018 (accessed August 21, 2020)}\natexlab{}.
\newblock \bibinfo{booktitle}{\emph{Kompleksowe Badania Ruchu we Wrocławiu i
  otoczeniu - KBR 2018}}.
\newblock
\urldef\tempurl%
\url{https://bip.um.wroc.pl/artykul/565/37499/kompleksowe-badania-ruchu-we-wroclawiu-i-otoczeniu-kbr-2018}
\showURL{%
\tempurl}


\end{thebibliography}

\end{document}